\documentstyle[11pt]{article}
\newcommand{\np}{Nucl. Phys. {\bf B}\ }
\newcommand{\pl}{Phys. Lett. {\bf B}\ }
\newcommand{\prd}{Phys. Rev. {\bf D}\ }

\newcommand{\prl}{Phys. Rev. Lett.\ }
\begin{document}
\begin{center}
{\large\bf
Symmetries at ultrahigh energies and searches for neutrino oscillations.}
\\
\vspace{.2in}
E. Papageorgiu,\\
LPTHE, Universit\'e de Paris XI, B\^at. 211, F-91405 Orsay.
\end{center}
\vspace{.5in}

\begin{abstract}
\noindent
Motivated by the possibility that new (gauge) symmetries which
are broken at the grand- (string-) unification scale give
rise to texture zeros in the fermion mass matrices which are
at the origin of the hierarchy of masses and mixings we explore
the effect of such zeros on the neutrino spectrum of SUSY-GUT models.
We find that the quadratic-seesaw spectrum on which most expectations
are focused is neither the only nor the most interesting possibility.
Cases of strong $\nu_{\mu} - \nu_{\tau}$ or $\nu_e - \nu_{\tau}$ mixing
are present for a specific texture structure of the Yukawa matrices
and experimental evidence can thus throw some light on the latter.
In contrast if the quadratic-seesaw scenario should be confirmed very
little could be said about the symmetries of the Yukawa sector.
\end{abstract}

\vskip 2 truecm
\noindent {\bf LPTHE Orsay Preprint 93-51}
\vskip 0.5cm
\noindent
August 1994
\newpage

\section{Introduction.}
One of the main deficiencies of the Standard Model is the lack of
understanding the structure of the Yukawa couplings $Y^{ij}$
in generation space $(i,j = 1,2,3)$.
In particular, the six quark and lepton masses plus the three  mixing
angles and the CP violating phase of the quark-mixing matrix
are not sufficient to fully specify the entries of the
mass matrices $M_{u,d,e} = Y_{u,d,e} <v_{u,d}> / \sqrt{2}$,
where the indices $u,d,e$ refer to the up- and down quarks and the
charged leptons.
In an attempt to keep the number of parameters which enter in the
latter minimal, various
authors have proposed (L-R) symmetric Yukawa matrices
containing two or three zero entries
\footnote{When counting the zeros of a symmetric matrix only the entries
above or below the diagonal are considered.},
so-called ``textures'', which, when multiplied by diagonal matrices
of phases, describe successfully the mass
spectra, and predict relations between the quark masses and the mixing
angles [1-4].
%\cite{{Fritzsch},{DHR},{Giudice},{RRR}}

Recently, the possibility of having L-R symmetric textures that can
correctly parametrise the up- and down-quark mass matrices with a
maximum number of five zeros at
the grand-unification scale $M_G \simeq 10^{16}$ GeV
has received special attention within the
context of the minimal supersymmetric standard model (MSSM) [2-7]
%\cite{{DHR},{DHRN},{Giudice},{RRR},{Leonta},{papageor}}
and a list of
five phenomenologically consistent solutions $(I)$, $(II)$, $(III)$,
$(IV)$ and $(V)$ were presented in
ref.\cite{RRR}. For convenience they have been parametrised in powers of
the Cabibbo angle $\lambda \simeq 0.22$ and the coefficients
$\alpha$, $\beta$, $\gamma$, $\delta$ and
$\alpha^{\prime}$, $\beta^{\prime}$, $\gamma^{\prime}$
whose values are given in Table 1:
\begin{equation}
Y_u = \left(
\begin{array}{ccc}
0 & \alpha \lambda^6 & \delta \lambda^4 \\
\alpha \lambda^6 & \beta \lambda^4 & \gamma \lambda^2 \\
\delta \lambda^4 & \gamma \lambda^2 & 1
\end{array}
\right) \,,
\end{equation}
and
\begin{equation}
Y_d = \left(
\begin{array}{rcl}
0 & \alpha^{\prime} \lambda^4 & 0 \\
\alpha^{\prime} \lambda^4 & \beta^{\prime} \lambda^3 & \gamma^{\prime}
\lambda^3 \\
0 & \gamma^{\prime} \lambda^3 & 1
\end{array}
\right) \,.
\end{equation}
%%%%%%%%%%%%%%%%%%%%%%%%%%%%%%%%%%%%%%%%%%%%%%%%%%%%%%%%%%%%%%%%%%%%%%%%

While a number of candidate models beyond the Standard Model (SM)
are (L-R) symmetric, -among others
some grand-unified (GUT) models that are based on the
$SO(10)$ gauge group-, the origin of the texture zeros which are responsible
for the well-known  mass-mixing relations in the quark sector
is not yet fully understood.
They are thought to be the relics of new fundamental
symmetries which are broken at or below the grand/string unification scale.
On the other hand, in L-R symmetric models it is natural to have
righthanded neutrino states $N_i$ and therefore Dirac neutrino mass terms:
$M_{\nu}^D N^c_i \nu_j$. The righthanded neutrinos acquire normaly
large Majorana masses
through radiative corrections [8-10]
or due to nonrenormalisable terms [11]
which originate from supergravity and string theory [12]:
\begin{equation}
M_{R} = {\cal C}\, {<H> <H>\over M_S}\, Y_R \,,
\end{equation}
where $M_S \sim 10^{18}$ GeV is the string unification scale and
the parameter ${\cal C}\sim 1 - 10^{-3}$ is characteristic of
large-radii orbifold compactification.
Since the Higgs field H acquires a
vacuum expectation value at $M_G$, typically the scale of the matrix $M_R$
lies in the intermediate mass range:
\begin{equation}
R \equiv {\cal C}\, {<H> <H>\over M_S}\, \simeq 10^{11} - 10^{14} {\rm GeV} \,.
\end{equation}
In ref.\cite{papageor} it was shown that in addition to the scale $R$,
also the texture structure of $Y_R$
plays a crucial role in the determination of the
mass spectrum of the  three superlight neutrino flavours: $\nu_e$, $\nu_{\mu}$,
$\nu_{\tau}$ and their mixing.

In particular, it is interesting to examine in greater detail the case where,
as a result of an extra symmetry at the Planck scale,
which may as well be at the origin of the perturbative structure
of the quark Yukawa matrices, some of the entries of the symmetric matrix
$M_R$, which for simplicity we assume to be real,
\begin{equation}
M_R = \left(
\begin{array}{ccc}
R_1 & R_4 & R_5 \\
R_4 & R_2 & R_6 \\
R_5 & R_6 & R_3
\end{array}
\right) \,,
\end{equation}
are zero while the others are of order R.
The underlying idea is that some of the higher order operators
of equ.(3) are forbidden when H becomes charged under an extra symmetry
which is not family blind.
On the other hand, there is evidence that
the presence of different powers of $\lambda$
at specific places in the quark Yukawa matrices,
is due to the breaking of such a $U(1)$ symmetry
through higher order operators containing heavy
Higgs fields and singlets [13].
One may therefore hope to find a common origin of the
mass- and mixing- hierarchy in the quark and lepton sector.

If one imposes that $M_R$ is a nonsingular matrix, the seesaw mechanism
guaranties
the existence of three light neutrinos, which are obtained upon diagonalisation
of the reduced mass matrix:
\begin{equation}
M_{\nu}^{eff} \simeq  M_{\nu}^{D\dag} M_R^{-1} M_{\nu}^{D}
\,.
\end{equation}
Then, if the connection of the lepton to the quark sector,
{\it i.e.} to the
Yukawa textures in eqs.(1,2), is made via the successful mass relations
of grand unification [14]:
\begin{equation}
M_{\nu}^{D} = M_u
\,,
\end{equation}
and
\begin{equation}
M_e = M_d
\,,
\end{equation}
where the (2,2) entry of $M_e$ has to be multiplied by a factor minus three
to account for discrepancies in the mass relations of the first two
generations,
it is possible to determine the neutrino spectrum for different
sets of textures and look for experimental tests.
As for the quark sector, we require the presence of
as many texture zeros in $M_R$ as this is compatible with its
nonsingularity. The corresponding four- and three-zero textures
are shown in Table 2.

In order to discuss the properties of the neutrino spectrum case by case,
it is usefull to write $M_{\nu}^{eff}$ in terms of the parameters
of eqs.(1-4).
The elements of the reduced matrix
\begin{equation}
M_{\nu_{ij}}^{eff} = {m_t^2 \over \Delta}\, m_{ij}
\end{equation}
with:
\begin{equation}
\begin{array}{l}
m_{11} = \delta^2 r_3 z^4 + \alpha^2 r_2 z^6 \\
m_{12} = m_{21} =  \gamma \delta r_3 z^3 + (\beta \delta + \alpha \gamma) r_6
z^4 + \alpha \beta r_2 z^5 + \alpha^2 r_4 z^6 \\
m_{13} = m_{31} = \delta r_3 z^2 + (\alpha + \gamma \delta) r_6 z^3
          + (\alpha \gamma r_2 + \delta^2 r_5) z^4 \\
m_{22}  =  \gamma ^2 r_3 z^2 + 2 \beta \gamma r_6 z^3 + \beta^2 r_2 z^4
           + 2 \alpha \gamma r_5 z^4 + 2 \alpha \beta r_4 z^5
           + \alpha^2 r_1 z^6 \\
m_{23}  = m_{32} = \gamma r_3 z + (\gamma^2 + \beta) r_6 z^2
           +(\beta \gamma r_2 + \alpha r_5 + \gamma \delta r_5) z^3
           +(\beta \delta + \alpha \gamma) r_4 z^4 \\
m_{33}  =  r_3 + 2 \gamma r_6 z + (\gamma^2 r_2 + 2 \delta r_5) z^2
           + 2 \gamma \delta r_4 z^3 + \delta^2 r_1 z^4 \,,
\end{array}
\end{equation}
are polynomials in $\lambda^2 \equiv z \simeq 0.05$
and  $\Delta \equiv det M_R$.
As in ref.\cite{papageor}, we denote by $r_{i = 1, ... 6}$ the minors
of the matrix $M_R$ which are obtained by omitting the row and column
containing the corresponding $R_i$ entry, {\it e.g.},
$r_3 = R_1 R_2 - R_4^2$.
When the minor  $r_3 \not= 0$ it sets the mass scale for the whole
matrix and the third-generation
neutrino:
\begin{equation}\label{tau}
m_{\nu_3} = {m_t^2\over R} \,,
\end{equation}
which becomes the hot dark-matter candidate, and the hierarchy
in the neutrino spectrum follows an analogous pattern as in the
quark sector. This implies that the heaviest (lightest) fermion belongs to
the third (first) generation and that the first-to-second generation mixing:
$|V_{\nu_e - \mu}| \sim \lambda/3$ prevails.
On the other hand, when $r_3 = 0$, {\it i.e.} for the textures
$M_R^{(b)}$ and
$M_R^{(d)}$ in Table 2, this natural pattern is in some cases
broken due to the appearance of a zero in the $m_{33}$
entry of $M_{\nu}^{eff}$ \cite{papageor}.

The mass eigenvalues of the matrix $M_{\nu}^{eff}$ can be determined
perturbatively from the characteristic equation:
\begin{equation}\label{P}
x^3 - r_3 f x^2 + z^4 r_1^{\star} g x - \Delta_{\nu} = 0 \,,
\end{equation}
where by $\Delta_{\nu}$ we have denoted the determinant of $m_{ij}$, and
$r_1^{\star} = r_2 r_3 - r_6^2$.
Then $f$ and $g$ are polynomials in $z$
which become of order one when all nonzero entries of $M_R$ are of order R:
\begin{eqnarray}\label{g}
f & = & 1 + 2 \gamma a_6 z + [(1 + a_2) \gamma^2 + 2 \delta a_5] z^2
        + 2 \gamma (\delta a_4 + \beta a_6) z^3 \nonumber \\
  &   & + [(1 + a_1) \delta^2 + \beta^2 a_2 + 2 \alpha \gamma a_5] z^4
        + 2 \alpha (\beta a_4 + \delta a_6) z^5 \nonumber \\
  &   &  + (a_1 + a_2) \alpha^2 z^6 \\
g & = & 1 - 2 a_1^{\star}
    (\alpha \gamma^2 - \alpha \beta + \beta \gamma \delta - 2 \gamma^3) z
\nonumber \\
  &   & + {\cal O}(z^2) + ... + {\cal O}(z^8) \nonumber \,,
\end{eqnarray}
with $a_i\equiv r_i/r_3$ and
$a_1^{\star} = (r_3 r_4 - r_5 r_6) /r_1^{\star}$.
Notice that $r_3 \not= 0$ holds for the textures $M_R^{(a)}$ and
$M_R^{(c)}$ while $r_1^{\star} \not= 0$ holds for the textures
$M_R^{(a)}$ and $M_R^{(d)}$, Table 2.

We start our discussion with the first Majorana texture $M_R^{(a)}$
for which $f$ and $g$ are functions of order one. Redefining next
$x \to x/R^2$ and using the fact that:
\begin{equation}
\begin{array}{c}
\Delta_{\nu} \simeq \kappa^2 z^{12} \cdot {\rm R}^6 \\
\kappa = \alpha^2 + \beta \delta^2 - 2 \alpha \gamma \delta \simeq {\cal O}
({\bf 1})\,,
\end{array}
\end{equation}
equ.(\ref{P}) reduces to:
\begin{equation}
x^3 - x^2 + z^4 x - z^{12} = 0 \,,
\end{equation}
which is the same for the five different types (I) - (V)
of the $Y_{u,d}$ textures,
given in eqs.(1,2) and Table 1.
One obtains an entirely model-independent neutrino-mass spectrum:
\begin{equation}\label{SP0}
m_{\nu_1}  \simeq {m_t^2\over R} z^8 \qquad
m_{\nu_2} \simeq {m_t^2\over R} z^4 \qquad
m_{\nu_3}  \simeq {m_t^2\over R} \,.
\end{equation}
The hierarchy implied by eq.(\ref{SP0}) is of the quadratic-seesaw type,
{\it i.e.}, the neutrino masses scale as the up-quark masses squared.

For the Majorana texture $M_R^{(c)}$ the neutrino mass spectrum is again
universal, {\it i.e.} independent of the particular structure of
quark Yukawa matrices.
In this case equ.(\ref{P}) reduces to:
\begin{equation}
x^3 - x^2 + z^5 x - z^{12} = 0 \,,
\end{equation}
and gives rise to a distorted seesaw spectrum:
\begin{equation}\label{H0}
m_{\nu_1} \simeq {m_t^2\over R} z^7 \qquad
m_{\nu_2} \simeq {m_t^2\over R} z^5 \qquad
m_{\nu_3} \simeq {m_t^2\over R}  \,.
\end{equation}

On the other hand for the remaining textures
$M_R^{(b)}$ and $M_R^{(d)}$ the universality is broken.
The characteristic equation assumes in general the form:
\begin{equation}
x^3 - z^n x^2 + z^m x - z^{12} = 0 \,,
\end{equation}
where $m = 4$ for $M_R^{(d)}$, and $m > 5$ for $M_R^{(b)}$,
while, $n = 1$ for $M_R^{(d)}$ and $Y_{u,d}^{(II),(IV),(V)}$,
$n = 2$ for $M_R^{(b)}$ and $Y_{u,d}^{(II)-(V)}$,
$n = 4$ for $M_R^{(d)}$ and $Y_{u,d}^{(III)}$ and for $M_R^{(b)}$ and
$Y_{u,d}^{(I)}$,
and $n = 6$ for $M_R^{(d)}$ and $Y_{u,d}^{(I)}$.
The resulting neutrino masses and mixings are shown in Table 3.
For simplicity we have assumed that there are no extra CP-violating
phases in the lepton sector so that the lepton-mixing matrix is given by:
\begin{equation}
V_l = U_{\nu} \, U_P \, U_e^{-1} \,,
\end{equation}
where $U_{\nu}$ and $U_e$ are the matrices diagonalising $M_{\nu}^{eff}$
and $M_e$ respectively while
\begin{equation}
U_P = \left(
\begin{array}{ccc}
1 & 0 & 0 \\
0 & 1 & 0 \\
0 & 0 & e^{i\phi}
\end{array}
\right) \qquad ,
\end{equation}
is the matrix relating the basis where $M_{\nu}^{eff}$ is diagonal to the basis
where $M_e$ is real.
 Written in powers of $\lambda$ and to lowest order,
\begin{equation}
U_e = \left(
\begin{array}{lcr}
1-\lambda^2/18 & -\lambda/3 & \gamma^{\prime} \lambda^4/3 \\
\lambda/3 & 1-\lambda^2/18 & -\gamma^{\prime} \lambda^3 \\
0 & \gamma^{\prime} \lambda^3 & 1
\end{array}
\right) \qquad .
\end{equation}

%@@@@@@@@@@@@@@@@@@@@@@@@@@@@@@@@@@@@@@@@@@@@@@@@@@@@@@@@@@@@@@@@@

The results of Table 3 confirm our previous statement that for the
texture $M_R^{(a)}$, which is almost proportional to the
unit matrix, and $M_R^{(c)}$ for which the minor $r_3$ is not zero
the usual hierarchy of fermion masses and mixing elements
is encountered also in the light-neutrino sector.
\footnote{
The only exception to this rule is found for the combined texture choice
$M_R^{(c)}$ with $Y_u^{(III)}$. In this case the lightest state is the
$\nu_{\mu}$ followed by the electron neutrino.}
The mass eigenstates $\nu_1,\,\nu_2,\,\nu_3$ obey the quadratic-seesaw
relation of eq.(16)
or the distorted seesaw relation of eq.(18), and the mixing between
the different neutrino flavours is given by:
\begin{equation}
sin^2 2\theta_{e-\mu} \simeq 0.02 \qquad
sin^2 2\theta_{\mu-\tau} \simeq 10^{-2} - 10^{-3} \qquad
sin^2 2\theta_{e-\tau} \simeq 2 \cdot 10^{-5} - 10^{-10} \,.
\end{equation}
Unfortunately, as long as the precise value of the entries in the
righthanded neutrino mass matrix is not specified these results
represent only an order of magnitude estimate.
In any case this most attractive scenario that can incorporate the
small-angle solution to the solar-neutrino problem through
$\nu_e \to \nu_{\mu}$ transitions and the tau neutrino as a candidate
of hot dark matter (HDM) will be for sure tested by the CHORUS and NOMAD
$\nu_{\mu} \to \nu_{\tau}$ oscillation experiments.
With respect to previous findings that a rather ``unsophisticated''
Majorana-mass sector, {\it i.e.}, one that contains no particular
symmetry or one that is proportional to the unity, would idealy lead to
a quadratic-seesaw scenario \cite{papageor}, it is interesting to add
the new cases involving mostly but not exclussively the texture
$M_R^{(c)}$.
Therefore if this scenario should receive experimental confirmation it
will become impossible to draw any conclusion on the structure of the
heavy Majorana sector.

Let us discuss next the more interesting cases of strong mixing that
follow from different mass patterns.
We find two cases of strong $\nu_{\mu} \to \nu_{\tau}$ mixing, both
containing the texture $M_R^{(d)}$, namely
$M_1 \sim Y_u^{(I)-1}  M_R^{(d)-1} Y_u^{(I)}$
and
$M_2 \sim Y_u^{(III)-1}  M_R^{(d)-1} Y_u^{(III)}$,
and two almost identical cases of strong $\nu_e \to \nu_{\tau}$
mixing where the texture $M_R^{(b)}$ is ivolved:
$M_3 \sim Y_u^{(II),(IV)-1}  M_R^{(b)-1} Y_u^{(II),(IV)}$.
The spectrum of $M_1$ and $M_2$ contains a very light $\nu_e$
and two mass-degenerate states which are linear combinations of
$\nu_{\mu}$ and $\nu_{\tau}$ with:
\begin{equation}
sin^2 2\theta_{e-\mu} \simeq 0.02 \qquad
sin^2 2\theta_{\mu-\tau} \simeq 1 \qquad
sin^2 2\theta_{e-\tau} \simeq 10^{-5} \,.
\end{equation}
A scenario with any of the standard neutrinos being a HDM
candidate is here rulled out but an explanation of the
atmospheric-neutrino deficit becomes an interesting possibility.
In the spectrum of $M_3$ the lightest (heaviest) state is a
$\nu_{\mu}$ ($\nu_{\tau}$) while $\nu_2$ is a linear combination
of $\nu_e$ and $\nu_{\tau}$ with:
\begin{equation}
sin^2 2\theta_{e-\mu} \simeq 0.02 \qquad
sin^2 2\theta_{\mu-\tau} \simeq 10^{-2} \qquad
sin^2 2\theta_{e-\tau} \simeq 1 \,.
\end{equation}
Here again the usual HDM scenario is excluded while the possibility
of explaining the solar-neutrino problem through a
$\nu_e \to \nu_{\tau}$ transition could be tested in the future.

In the cases,
$M_4 \sim Y_u^{(III)-1}  M_R^{(c)-1} Y_u^{(III)}$
and
$M_5 \sim Y_u^{(IV)-1}  M_R^{(d)-1} Y_u^{(IV)}$,
the hierarchy between $\nu_e$ and $\nu_{\mu}$ is flipped
with respect to the usual seesaw spectrum but without alteration
of the mixing angles in eq.(23).
The case $M_6 \sim Y_u^{(I)-1}  M_R^{(b)-1} Y_u^{(I)}$ is an
interesting example of a scenario with $\nu_{\mu}$ as the
heaviest neutrino \cite{papageor} (while $\nu_{\tau}$ is the
second heaviest state) and vanishing
$\nu_e \to \nu_{\tau}$ mixing.
For the $\nu_{\mu}$ to become a HDM candidate the Majorana mass scale
should be: $R \sim z^3\times 10^{12}$ GeV which is typical of
radiatively generated righthanded neutrino masses in nonsupersymmetric GUTs.

The cases involving $Y_{u,d}^{(V)}$ are treated separately due to the
remarkable stability that the neutrino spectrum exhibits as a function
of the structure of the Majorana mass matrix. This is due to the fact
that the $m_{ij}$ entries of the light neutrino mass matrix, equ.(10),
are on one hand power series of $z$
with coefficients $\beta, \gamma, \delta$ (up to $z^4$) which
are of order one for the up-quark texture $Y_u^{(V)}$, on the other
hand they are ordered series of the minors $r_3; r_6; r_2,r_5; r_4;
r_1$. Therefore the hierarchical structure of $M_{\nu}^{eff}$
up to fourth order in $z$ is, independently of $M_R$,
\begin{equation}
M_{\nu}^{(V)} \sim {m_t^2 z^p\over R} \left(
\begin{array}{ccc}
z^4 & z^3 & z^2 \\
z^3 & z^2 & z \\
z^2 & z & 1
\end{array}
\right) \qquad ,
\end{equation}
where the power $p$ can in principle go from zero to four.
In all models of this type one has a quadratic-seesaw mass spectrum,
but for which the mixing between $\nu_e$ and
$\nu_{\mu}$ turns out to be two to three times larger than usual:
\begin{equation}
sin^2 2\theta_{e-\mu} \simeq 0.05 \,,
\end{equation}
the mixing between $\nu_e$ and $\nu_{\tau}$ exceeds by far the seesaw
expectations:
\begin{equation}
sin^2 2\theta_{e-\tau} \simeq 0.9\cdot 10^{-2} \,,
\end{equation}
while the mixing between $\nu_{\mu}$ and $\nu_{\tau}$ is negligible.
This anomaly is representative of the structure of the quark-Yukawa
sector and has been discussed in ref.(\cite{papageor}).

\vskip 1cm
\noindent
{\bf Acknowledgements:}
This work was supported by EEC Project: ERBCHBICT930850.

\newpage
%%%%%%%%%%%%%%%%%%%%%%%%%%%%%%%%%%%%%%%%%%%%%%%%%%%%%%%%%%%%%%%%%%%%%%%%
\vskip 1cm
{\bf Table 1:} The values of the parameters in equ.(45) and equ.(46)
that correspond to the five distinct classes of maximally-predictive
GUT models from ref.[7].
\vskip 1cm
\begin{tabular}{|l|l|l|l|l|l|} \hline
& & & & & \\
& {\bf (I)} & {\bf (II)} & {\bf (III)} & {\bf (IV)} & {\bf (V)} \\
& & & & & \\
\hline
{\bf $\alpha$} & $\sqrt{2}$ & 1 & 0 & $\sqrt{2}$ & 0 \\
\hline
{\bf $\beta$} & 1 & 0 & 1 & $\sqrt{3}$ & $\sqrt{2}$ \\
\hline
{\bf $\gamma$} & 0 & 1 & 0 & 1 & 1/$\sqrt{2}$ \\
\hline
{\bf $\delta$} & 0 & 0 & $\sqrt{2}$ & 0 & 1 \\
\hline
$\alpha^{\prime}$ & 2 & 2 & 2 & 2 & 2 \\
\hline
$\beta^{\prime}$ & 2 & 2 & 2 & 2 & 2 \\
\hline
{\bf $\gamma^{\prime}$} & 4 & 2 & 4 & 0 & 0 \\
\hline
\end{tabular}
\vskip 1cm
%%%%%%%%%%%%%%%%%%%%%%%%%%%%%%%%%%%%%%%%%%%%%%%%%%%%%%%%%%%%%%%%%%%%%
\vskip 1cm
\noindent
{\bf Table 2:} Nonsingular symmetric textures with a maximum of zero entries
representing the Majorana mass matrix $M_R$ of the righthanded neutrinos.
\vskip 0.5 cm
$
M_R^{(a)} = \pmatrix{
R_1 &0 &0 \cr
0 &R_2 &0 \cr
0 &0 &R_3 \cr
}$
\hskip 4 truecm
$M_R^{(b)} = \pmatrix{
0 &0 &R_5 \cr
0 &R_2 &0 \cr
R_5 &0 &0 \cr
}$
\par \bigskip

$M_R^{(c)} = \pmatrix{
0 &R_4 &0 \cr
R_4 &0 &0 \cr
0 &0 &R_3 \cr
}$
\hskip 4 truecm
$M_R^{(d)} = \pmatrix{
R_1 &0 &0 \cr
0 &0 &R_6 \cr
0 &R_6 &0 \cr
}$ \par \bigskip

\newpage
%%%%%%%%%%%%%%%%%%%%%%%%%%%%%%%%%%%%%%%%%%%%%%%%%%%%%%%%%%%%%%%%%%%%%
{\bf Table 3:}  The neutrino spectrum for the four righthanded-neutrino
textures $M_R^{(a)-(d)}$ of Table 2 and the different parametrisations
of the quark-Yukawa matrices $Y_{u,d}^{(I)-(IV)}$ defined by the
eqs.(1,2) and Table 1.

\vskip 3 mm
\vfill \eject
\begin{table}
\begin{tabular}{cccccccc}
$M_{\nu}^{eff}\, \simeq$ & ${<v> \over \sqrt{2}}\,\, Y_u^{(I)^T}
\left [ \right.$ & $M_R^{(a)_{-1}}\,;$ & $M_R^{(b)_{-1}}\,;$ &
$M_R^{(c)_{-1}}\,;$
& $M_R^{(d)_{-1}}$ & $\left. \right ] $ & $Y_u^{(I)}$ \cr
& & & & & & &  \cr
& & & & & & & \cr
$m_{\nu_1}$ & \qquad$\simeq$ & $z^8$ & $z^6$ & $z^7$ & $z^8$ &
$\times \, {m_t^2\over R}$ & \cr
& & & & & & & \cr
$m_{\nu_2}$ & \qquad$\simeq$ & $z^4$ & $z^3$ & $z^5$ & $z^2$ &
$\times \, {m_t^2\over R}$  & \cr
& & & & & & & \cr
$m_{\nu_3}$ & \qquad$\simeq$ & 1 & $z^3$ & 1 & $z^2$ &
$\times \, {m_t^2\over R}$ & \cr
& & & & & & & \cr
$|V_{\nu_1-\mu}|$ & \qquad$\simeq$ & ${\lambda\over 3}$
& ${\lambda\over 3}$ & ${\lambda\over 3}$ & ${\lambda\over 3}$ & & \cr
& & & & & & & \cr
$|V_{\nu_2 - \tau}|$ & \qquad$\simeq$ & $4\lambda^3$ & 1
& $4\lambda^3$ & 1 & & \cr
& & & & & & & \cr
$|V_{\nu_1 - \tau}|$ & \qquad$\simeq$ & 0 & 0 & $4\lambda^5$
& $4\lambda^5$ & & \cr
\end{tabular}
\end{table}

\vskip 3 mm
\begin{table}
\begin{tabular}{cccccccc}
$M_{\nu}^{eff}\, \simeq$ & ${<v> \over \sqrt{2}}\,\, Y_u^{(II)^T}
\left [ \right.$ & $M_R^{(a)_{-1}}\,;$ & $M_R^{(b)_{-1}}\,;$ &
$M_R^{(c)_{-1}}\,;$
& $M_R^{(d)_{-1}}$ & $\left. \right ] $ & $Y_u^{(II)}$ \cr
& & & & & & &  \cr
& & & & & & & \cr
$m_{\nu_1}$ & \qquad$\simeq$ & $z^8$ & $z^6$ & $z^7$ & $z^8$ &
$\times \, {m_t^2\over R}$ & \cr
& & & & & & & \cr
$m_{\nu_2}$ & \qquad$\simeq$ & $z^4$ & $z^4$ & $z^5$ & $z^3$ &
$\times \, {m_t^2\over R}$  & \cr
& & & & & & & \cr
$m_{\nu_3}$ & \qquad$\simeq$ & 1 & $z^2$ & 1 & $z$ &
$\times \, {m_t^2\over R}$ & \cr
& & & & & & & \cr
$|V_{\nu_1-\mu}|$ & \qquad$\simeq$ & ${\lambda\over 3}$
& 1 & ${\lambda\over 3}$ & ${\lambda\over 3}$ & & \cr
& & & & & & & \cr
$|V_{\nu_2 - \tau}|$ & \qquad$\simeq$ & $\lambda^2$ & 1
& $\lambda^2$ & $\lambda^2$ & & \cr
& & & & & & & \cr
$|V_{\nu_1 - \tau}|$ & \qquad$\simeq$ & $\lambda^4$ & $\lambda^2$
& $\lambda^4$ & $\lambda^4$ & & \cr
\end{tabular}
\end{table}

\vskip 5 mm
\begin{table}
\begin{tabular}{cccccccc}
$M_{\nu}^{eff}\, \simeq$ & ${<v> \over \sqrt{2}}\,\, Y_u^{(III)^T}
\left [ \right.$ & $M_R^{(a)_{-1}}\,;$ & $M_R^{(b)_{-1}}\,;$ &
$M_R^{(c)_{-1}}\,;$
& $M_R^{(d)_{-1}}$ & $\left. \right ] $ & $Y_u^{(III)}$ \cr
& & & & & & &  \cr
& & & & & & & \cr
$m_{\nu_1}$ & \qquad$\simeq$ & $z^8$ & $z^6$ & $z^7$ & $z^8$ &
$\times \, {m_t^2\over R}$ & \cr
& & & & & & & \cr
$m_{\nu_2}$ & \qquad$\simeq$ & $z^4$ & $z^4$ & $z^5$ & $z^2$ &
$\times \, {m_t^2\over R}$  & \cr
& & & & & & & \cr
$m_{\nu_3}$ & \qquad$\simeq$ & 1 & $z^2$ & 1 & $z^2$ &
$\times \, {m_t^2\over R}$ & \cr
& & & & & & & \cr
$|V_{\nu_1-\mu}|$ & \qquad$\simeq$ & ${\lambda\over 3}$
& ${\lambda\over 3}$ & 1 & ${\lambda\over 3}$ & & \cr
& & & & & & & \cr
$|V_{\nu_2 - \tau}|$ & \qquad$\simeq$ & $4\lambda^3$ & $4\lambda^3$
& $\lambda^4$ & 1 & & \cr
& & & & & & & \cr
$|V_{\nu_1 - \tau}|$ & \qquad$\simeq$ & $\lambda^4$ & $\lambda^4$
& $4\lambda^3$ & $\lambda^4$ & & \cr
\end{tabular}
\end{table}

\vskip 5 mm
\begin{table}
\begin{tabular}{cccccccc}
$M_{\nu}^{eff}\, \simeq$ & ${<v> \over \sqrt{2}}\,\, Y_u^{(IV)^T}
\left [ \right.$ & $M_R^{(a)_{-1}}\,;$ & $M_R^{(b)_{-1}}\,;$ &
$M_R^{(c)_{-1}}\,;$
& $M_R^{(d)_{-1}}$ & $\left. \right ] $ & $Y_u^{(IV)}$ \cr
& & & & & & &  \cr
& & & & & & & \cr
$m_{\nu_1}$ & \qquad$\simeq$ & $z^8$ & $z^6$ & $z^7$ & $z^8$ &
$\times \, {m_t^2\over R}$ & \cr
& & & & & & & \cr
$m_{\nu_2}$ & \qquad$\simeq$ & $z^4$ & $z^4$ & $z^5$ & $z^3$ &
$\times \, {m_t^2\over R}$  & \cr
& & & & & & & \cr
$m_{\nu_3}$ & \qquad$\simeq$ & 1 & $z^2$ & 1 & $z$ &
$\times \, {m_t^2\over R}$ & \cr
& & & & & & & \cr
$|V_{\nu_1-\mu}|$ & \qquad$\simeq$ & ${\lambda\over 3}$
{}~~~~~& 1 & ${\lambda\over 3}$ & 1 & & \cr
& & & & & & & \cr
$|V_{\nu_2 - \tau}|$ & \qquad$\simeq$ & $\lambda^2$ & 1
& $\lambda^2$ & $\lambda^4$ & & \cr
& & & & & & & \cr
$|V_{\nu_1 - \tau}|$ & \qquad$\simeq$ & $\lambda^4$ & $\lambda^2$
& $\lambda^4$ & $\lambda^2$ & & \cr
\end{tabular}
\end{table}

%%%%%%%%%%%%%%%%%%%%%%%%%%%%%%%%%%%%%%%%%%%%%%%%%%%%%%%%%%%%%%%%%%%%%%%%

\end{document}